\documentclass[12pt, journal, onecolumn]{IEEEtran}
\date{}

\usepackage{subfigure}
\usepackage{amsmath,amssymb}
\usepackage{color}
\usepackage{graphicx}
\usepackage{setspace}

\newtheorem{theorem}{Theorem}
\newtheorem{proposition}{Proposition}
\newtheorem{lemma}{Lemma}

\newtheorem{remark}{Remark}
\allowdisplaybreaks

\begin{document}
%
\title{Power Allocation for Energy Harvesting Transmitter with Causal Information}
%
%
%

\author{Zhe~Wang,
        Vaneet~Aggarwal,
        Xiaodong~Wang
\thanks{Z. Wang and X. Wang are with the Electrical Engineering Department, Columbia University, New York, NY 10027 (e-mail:
\{zhewang, wangx\}@ee.columbia.edu).}
\thanks{V. Aggarwal is with AT\&T Labs-Research, Bedminster, NJ 07921 USA (e-mail: vaneet@research.att.com).}}

\maketitle

\begin{abstract}
We consider power allocation for an access-controlled transmitter with energy harvesting capability based on causal observations of the channel fading state. We assume that the system operates in a time-slotted fashion and the channel gain in each slot is a random variable which is independent across slots. Further, we assume that the transmitter is solely powered by a renewable energy source and the energy harvesting process can practically be  predicted. With the additional access control for the transmitter and the maximum power constraint, we formulate the stochastic optimization problem of maximizing the achievable rate as a Markov decision process (MDP) with continuous state. To efficiently solve the problem, we define an approximate value function based on a piecewise linear fit in terms of the battery state. We show that with the approximate value function, the update in each iteration consists of a group of convex problems with a continuous parameter.  Moreover, we derive the optimal solution to these convex problems in closed-form. Further, we propose power allocation algorithms  for both the finite- and infinite-horizon cases, whose computational complexity is significantly lower than that of the standard discrete MDP method but with improved performance. Extension to the case of a general payoff function and imperfect energy prediction is also considered. Finally, simulation results demonstrate that the proposed algorithms closely approach the optimal performance.
\end{abstract}

\begin{IEEEkeywords}
Causal information, energy harvesting, fading channel, Markov decision process, power allocation.
\end{IEEEkeywords}

\newpage

\section{Introduction}

The utilization of renewable energy is an important characteristic of the green wireless communication systems~\cite{GreenRadio}. Renewable energy powered transmitters  can be deployed for  wireless sensor networks or cellular networks, reducing the reliance on traditional batteries and prolonging the transmitter's lifetime~\cite{FundamentalTradeoff}\cite{Green4G}. However, the fluctuation of the energy harvesting together with the variation of the channel fading brings many challenges to the design of energy-harvesting  communication systems~\cite{EnergyHarvesting}\cite{ASurvey}.

Wireless transmission schemes for energy-harvesting transmitters have been investigated by a number of recent works~\cite{FiniteHorizon}\cite{OptimalEnergy}\cite{TransmissionEnergy}\cite{MyPaper}. In order to achieve the optimal throughput, a ``shortest path" based energy scheduling algorithm was proposed in \cite{FiniteHorizon} for a static channel with finite battery capacity and non-causal energy harvesting state. The authors of \cite{OptimalEnergy} discussed an MDP model for the case when the energy harvesting and channel fading are known causally and there is no maximum power constraint. A staircase water-filling algorithm was proposed in \cite{OptimalEnergy} for the case when the battery capacity is  infinite, and the  energy harvesting and fading channel states are known non-causally. With a finite battery capacity  and non-causal energy harvesting and fading channel states, a  water-filling procedure was studied in \cite{TransmissionEnergy}, and with an additional maximum power constraint a  dynamic water-filling algorithm was proposed  in \cite{MyPaper}. The authors of \cite{Neely} developed an online approximately optimal algorithm based on Lyapunov optimization, which is designed to maximize a utility function, based on the number of packet transmissions in energy harvesting networks. In \cite{LearningTh}, using the  discrete MDP model, a reinforcement learning based approach was used to optimize the number of packet transmissions without the prior knowledge of the statistics of the energy harvesting process and the channel fading process. The authors of \cite{DiscretePI} considered a static channel with causal knowledge of the stationary Poisson energy arrival process and gave an MDP-based solution to maximize the average throughput with unconstrained transmission power. On the other hand, the throughput optimization problem with causal information on the energy harvesting state and the fading channel state, and under the maximum power constraint, remains open.  In this paper, we will tackle this problem.

Specifically, we first consider the power allocation for an access-controlled transmitter, which is powered by a renewable energy source and equipped with a finite-capacity battery and has a maximum power constraint. The channel fading is assumed to be a random variable in a slot and is independent across different slots. For energy harvesting, we first assume that it can be predicted accurately for the scheduling period, which can be realized in practice \cite{PMEHW}\cite{AMPEAE}, and then later introduce the prediction error variables. Furthermore, we assume that a control center can temporarily suspend the transmitter's access due to channel congestion. Such channel access control for the transmitter is modeled as a first-order Markov process. Under the above setting, this paper finds the approximately optimal power allocation for both the finite- and infinite-horizon cases.

To obtain the power allocation, we formulate the stochastic optimization problem as a discrete-time and continuous-state Markov decision process (MDP), with the objective of maximizing the sum of the payoff in the current slot and the discounted expected  payoffs in the future slots, where the payoff function is the achievable channel rate.  Since the state variables including the battery state and the channel state in the MDP problem are continuous, to avoid the prohibitively high complexity for updating the value function caused by the continuous states, this paper introduces an approximate value function. We show that the approximate value function is concave and non-decreasing in the variable corresponding to the energy stored in the battery,  which further enables the approximate value function be updated  in closed-form. This is then used to find the approximately optimal solution of the power allocation for both the finite- and  infinite-horizon cases.
%

The proposed algorithms provide approximate solutions, whose performances are lower bounded by the standard discrete MDP method. Also, to obtain the solution, we solve at most ${\cal O}(B_{\max}/\delta \cdot C)$ convex optimization problems where $B_{\max}$ is the battery capacity,  $\delta$ is the approximation precision,  and $C$ is the length of horizon for the finite-horizon case or the maximum number of iterations for the infinite-horizon case. In particular, for the infinite-horizon case, given a convergence tolerance $\alpha$, the $\alpha$-converged solution can be obtained within ${\cal O}(\log_\gamma\alpha)$ iterations, where $\gamma$ is the discount factor.

The remainder of the paper is organized as follows. In Section II, we describe the system model, formulate the energy scheduling problem as a continuous-state MDP problem and define the value function. In Section III, we define an approximate value function and prove that the approximate value function is non-decreasing and concave with respect to the continuous battery state.  In Section IV, we derive the optimal closed-form procedure for updating the approximate value function and develop the power allocation algorithms for both finite- and infinite-horizon cases. The proposed algorithms are extended to deal with the model with a general payoff function and imperfect energy prediction in Section V. Section VI provides simulation results and Section VII concludes the paper.

\section{Problem Formulation}
\subsection{System Model}

\begin{figure}
\centering
\includegraphics[width=0.7\textwidth]{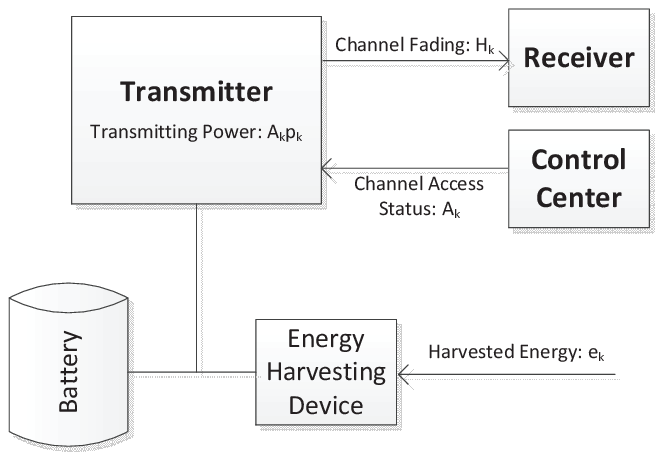}
\caption{The system block diagram.}
\label{fg:system}
\end{figure}

We consider a point-to-point communication system with one transmitter and one receiver, as shown in Fig.~\ref{fg:system}. We assume a slow fading channel model where the channel gain is constant for a coherence time of $T_c$ (corresponding to a time slot) and changes independently across slots. The signal model for slot $k$ is given by
\begin{equation}
\boldsymbol{y}_k = H_k \boldsymbol{x}_k + \boldsymbol{w}_k,
\end{equation}
where $\boldsymbol{y}_k\in \mathbb{C}^{T_c}$ is the received signal, $\boldsymbol{x}_k \in \mathbb{C}^{T_c}$ is the transmitted signal, $H_k \in \mathbb{C}$ is the channel gain in slot $k$ and $\boldsymbol{w}_k \in \mathbb{C}^{T_c}$ is the additive white Gaussian noise consisting of $\mathbb{CN}(0,1)$ elements.

At the beginning of each slot, the transmitter is informed of the channel access status $A_k\in\{0,1\}$ for the current slot from the control center, where $A_k=0$ indicates that the channel access is not permitted for slot $k$ while $A_k=1$ indicates otherwise. We assume that $A_k$ follows a stationary first-order Markov process, whose transition probabilities are given as   $\textrm{Pr}(A_{k+1}=0\;|\;A_{k}=1)=q_k$ and $\textrm{Pr}(A_{k+1}=0\;|\;A_{k}=0)=\tilde{q}_k$.  If $A_k=0$, then the transmit power in slot $k$ is $p_k=0$. On the other hand, if $A_k=1$, then the transmitter needs to decide its transmit power $p_k$.

The transmitter is powered by an energy harvesting device, e.g., a solar panel, and a battery. The battery, which buffers the harvested energy, has a finite capacity, denoted by $b_{\max}$. Since the energy harvesting process is steady or can be well predicted, we assume that the energy harvested over the next $K$ slots can be non-causally known, denoted as $e_k$ (the causal energy harvesting model will be considered in Section V). We assume $h_k\triangleq |H_k|^2$ is independent across slots (i.i.d. when $K=\infty$).

In slot $k$, the transmitter transmits at a power level of  $p_k$ ($p_k=0$ if $A_k=0$), which is constrained by the maximum transmission power $p_{\max}$ and the available energy $b_k$, i.e.,
\begin{equation}\label{eq:powerc}
0\leq p_k \leq \min\big\{p_{\max},b_k/T_c\big\}\ .
\end{equation}
The battery level at the beginning of slot $k+1$ is given as
\begin{equation}\label{eq:battery}
b_{k+1} = \min\big\{b_{\max}, b_k  + e_k - p_kT_c\big\} \ ,
\end{equation}
with the constraint that the  battery level is non-negative for all slots, i.e.,
\begin{equation} \label{eq:batteryc}
b_k\geq 0\ .
\end{equation}
Further, the transmitter receives a  payoff $r(p,h)$ based on the transmission power and channel gain. In this paper, we use the achievable channel rate as the payoff, i.e., $r(p, h) = \log(1+ph)$. Also, in Section V, we consider a general payoff function $r(p,h)$ which is continuous, non-decreasing, and concave with respect to  $p$ given $h$.

\subsection{Problem Formulation}
We assume that $e_k$ can be predicted non-causally while all other variables are only known causally to the transmitter (we will relax this assumption in Section V where we assume that $e_k$ is predicted with a random error $\varepsilon_k$).  Denote $\boldsymbol{H} \triangleq [h_1,h_2,\ldots,h_K]$,  $\boldsymbol{A} \triangleq [A_1,A_2,\ldots,A_K]$, and a discount factor $\gamma\in[0,1]$. We assume that all the side information, e.g., the distributions of all random variables and the predictions of the harvested energy, is known before the first slot. Then the power allocation policy ${\cal P}\triangleq\{p_k(\Gamma_k)\;|\;k=1,2,\ldots,K\}$ needs to be calculated to maximize the expected total payoff in the next $K$ slots, where $\Gamma_k\triangleq(b_k,h_k,A_k)$ consists of the observations available at the beginning of slot $k$. Since $b_k$ and $h_k$ are continuous variables, it is not possible  to store ${\cal P}$ in a look-up table. Instead, we only store some of the intermediate results, i.e., the approximate value function introduced in Section III, in an efficient way, and then calculate the power allocation when $\Gamma_k$ is observed. Specifically, at the beginning of slot $k$, given $\Gamma_k$, if channel access is permitted, i.e., $A_k=1$,  the transmitter calculate the power level $p_k$. And if the channel access is not permitted, i.e., $A_k=0$, then $p_k=0$. To that end, we formulate the following optimization problem for defining the optimal policy
\begin{equation}\label{eq:problem}
{\cal P}^* \triangleq \arg \max_{p_k(\cdot),k=1,2,\ldots,K}\Big\{\mathbb E_{\boldsymbol{H},\boldsymbol{A}} \Big[\sum_{k=1}^K \gamma^{k-1} \log(1+ p_k(\Gamma_k)h_k)\Big]\Big\}\ ,
\end{equation}
subject to the constraints in \eqref{eq:powerc}, \eqref{eq:battery}, and \eqref{eq:batteryc} for $k = 1,2,\ldots,K$.

Note that by \eqref{eq:battery}, the battery level $b_k$ forms a continuous-state first-order Markov chain, whereas the channel access state $A_k$ is a discrete-state Markov chain  by assumption. Then, we can convert the problem in \eqref{eq:problem} to its equivalent MDP recursive form \cite{MDP} in terms of the {\em value function}, which represents the total payoff received in the current slot and expected to be received in the future slots.

Specifically, in the MDP model we treat the battery level $b$ and the channel access state $A$, i.e., $(b,A)$,  as the state, the channel $h$ as the observation, and the transmit power $p$ as the decision. Then, the state space becomes $\{0\leq b\leq b_{\max}\}\times\{0,1\}$; and the corresponding decision space is ${\cal D}_1(b)=\{0\leq p \leq \min\{b/T_c, p_{\max}\}\}$ and  ${\cal D}_0=\{0\}$, corresponding to  $A=1$ and $A=0$, respectively. The value function is then recursively defined as
\begin{equation}
v^k(b_k,A_k)\triangleq \mathbb{E}_{h_k}\Big[\max_{p_k(\Gamma_k)\in{\cal D}_{A_k}(b_k)} \big\{ \log(1 + p_k(\Gamma_k)h_k) + \gamma  u ^{k}(b_k,p_k(\Gamma_k),A_k)\big\}\Big]\ , k=1,2,\ldots,K\ ,\label{eq:blm2}
\end{equation}
where
\begin{align}
 u^{k}(b_k,p_k,A_k) &\triangleq \mathbb{E}_{A_{k+1}|A_k}\big[v^{k+1}(\min\{b_{\max},b_k + e_k - p_kT_c\},A_{k+1}) \big]\label{eq:blm3}\ ,
\end{align}
and
\begin{align}
v^{K+1}(b,A)=0, \ \textrm{for all } b\in[0,b_{\max}], A\in\{0,1\}\ .
\end{align}
Note that, $v^k(b_k,A_k)$ represents the expected maximum discounted  payoff between slots $k$ and $K$ given the side information $b_k$ and $A_k$. Due to the causality and the backward recursion, the observation $\Gamma_k$ in slot $k$ does not affect the value function for slot $k+1$. Also, when $A_k=1$, given the value function for slot $k+1$, the optimal power allocation for slot $k$ can be obtained by
\begin{align}\label{eq:omdpeq}
p_k^*(\Gamma_k) = \arg \max_{p\in{\cal D}_{A_k}(b_k)}\big\{& \log(1+ph_k) + \gamma  u ^k(b_k,p,1)\big\}\ ,
\end{align}
where $u ^k(b,p,A)$ is calculated using \eqref{eq:blm3}. Moreover, when $A_k=0$, we always have
\begin{equation}
p_k^*(\Gamma_k)= 0\ .
\end{equation}

\section{Approximate Value Function}
By recursively computing the value function $v^k(b,A)$ defined in \eqref{eq:blm2}, in theory we can obtain the optimal solution to  \eqref{eq:omdpeq} for each $k\in\{1,2,\ldots,K\}$. However, a closed-form expression for $v^k(b,A)$ is hard to obtain when $K$ is large, e.g., $K\geq 3$. A typical approach is to quantize the continuous variables $(b,p,h)$ to finite number of discrete levels, i.e., to convert the original problem to a discrete MDP problem \cite{MDP}. However, with such discretization, solving the corresponding discrete MDP problem involves an exhaustive search on ${\cal D}_1(b)$ for all discretized $h$, and we can only obtain discrete power levels.

In order to efficiently solve the MDP problem and obtain the continuous power allocation, in this section, we will define an approximate value function by using a piecewise linear approximation based on some discrete samples of $\{v^k(B,A)\;|\;\ B\in\{0,\delta,2\delta,\ldots,b_{\max}\},A\in\{0,1\}\}$ where $\delta$ is an approximation precision. This approximate value function is shown to be concave and non-decreasing  in the variable corresponding to the energy stored in the battery, making the optimal power allocation problem in \eqref{eq:omdpeq} (or \eqref{eq:mdpeq}) a convex optimization problem.

\subsection{ Value Function Approximation}
With an approximation precision  parameter $\delta$, we define a piecewise linear approximation operator:
\begin{equation}\label{eq:oavf}
{\cal L}\left[v^k(b,A),\delta\right] \triangleq v^k(\lfloor b/ \delta\rfloor\delta,A) + \frac{b-\lfloor b/\delta \rfloor \delta} {\delta}\big(v^k(\lceil b/\delta\rceil \delta,A)-v^k(\lfloor b/\delta\rfloor \delta,A)\big)
,\  b\in[0,b_{\max}]\ ,
\end{equation}
and ${\cal L}\left[v^K(b,A),\delta\right]\triangleq v(b_{\max},A)$ for any $b>b_{\max}$, as shown in Fig. \ref{fg:bd}.

Initially, we define
\begin{equation}\label{eq:apinit}
W_\delta^K(b,A)\triangleq {\cal L}\left[v^K(b,A),\delta\right]\ ,
\end{equation}
which is a linear approximation to $v^K(b,A)$.
Then, recursively from $k=K-1$ to $k=1$, we use the approximate value function to replace the original value function in \eqref{eq:blm3}, i.e., $v^k(b,A)\leftarrow W^k_\delta(b,A)$,  and define
\begin{equation}\label{eq:nblm3}
 U^{k}(b_k,p_k,A_k) \triangleq \mathbb{E}_{A_{k+1}|A_k}\big[W_{\delta}^{k+1}(\min\{b_{\max},b_k + e_k - p_kT_c\},A_{k+1}) \big]\ .
\end{equation}
By setting $u^k(b_k,p_k,A_k)\leftarrow  U^{k}(b_k,p_k,A_k)$ in \eqref{eq:blm2}, we further define
\begin{equation}\label{eq:nblm2}
V^k(b_k,A_k)\triangleq \mathbb{E}_{h_k}\Big[\max_{p_k(\Gamma_k)\in{\cal D}_{A_k}(b_k)} \big\{ \log(1+p_k(\Gamma_k)h_k) + \gamma  U^{k}(b_k,p_k(\Gamma_k),A_k)\big\}\Big]\ .
\end{equation}
Finally, we write the approximation value function as
\begin{equation}\label{eq:avf}
W_\delta^k(b,A)\triangleq {\cal L}\left[V^k(b,A),\delta\right]\ .
\end{equation}
Note that, in \eqref{eq:nblm3}-\eqref{eq:avf}, we made the substitutions $v^k(b,A)\leftarrow W^k_\delta(b,A)$ and  $u^k(b_k,p_k,A_k)$ in \eqref{eq:blm3} and \eqref{eq:blm2}, respectively. Thus we can treat the approximate value function $W_{\delta}^k(b,A)\triangleq {\cal L}\left[V^k(b,A),\delta\right]$, which is updated by \eqref{eq:nblm3}-\eqref{eq:avf}, as an approximation to the value function $v^k(b,A)$, which is updated by \eqref{eq:blm2}-\eqref{eq:blm3}.

We consider the approximation error $||W_{\delta}^k(b,A)-v^k(b,A)||_\infty$ at slot $k$ (or iteration $i=K-k+1$). In each iteration, the error is produced by the piecewise linear approximation in \eqref{eq:avf} and propagated through solving the problem in \eqref{eq:nblm2}. Then, at the end of each iteration the total error accumulated by the obtained approximate value function is the sum of the newly produced error and the discounted propagated error, growing with the iteration number. Since the update rules for both $v^k(b,A)$ and $W_{\delta}^k(b,A)$ start from the same initial value function $v^K(b,A)$, then the total error in the $i$-th iteration  (we use the subscript $(i)$ to denote the $i$-th iteration, which represents slot $K-i+1$) can be bounded by
\begin{equation}\label{eq:bound}
||W^{(i)}_{\delta}(b,A)-v^{(i)}(b,A)||_{\max} \leq \sum_{j=1}^i \gamma^{i-j}\epsilon_j(\delta)
\end{equation}
where
\begin{equation}\label{eq:err}
 \epsilon_j(\delta) \triangleq \max_{b\in[0,b_{\max}],A\in\{0,1\}}\{V^{(j)}(b,A) - W_{\delta}^{(j)}(b,A)\} = ||V^{(j)}(b,A) - W_{\delta}^{(j)}(b,A)||_{\infty}
\end{equation}
is the new error produced by \eqref{eq:avf} in the $j$-th iteration.

With the approximate value function for each slot $k$, when $A = 1$, the power allocation given $\Gamma$ can be obtained by
\begin{equation}\label{eq:mdpeq}
p_k^*(\Gamma) = \arg \max_{p\in{\cal D}_{1}(b)}\big\{ \log(1+ph) + \gamma  U^k(b,p,1)\big\}\ .
\end{equation}

Define ${\cal B}_{\delta}\triangleq \{0,\delta,2\delta,\ldots,b_{\max}\}$. Note that the approximate value function is linearly recovered from the sample set $\{V^k(b,A)\;|\;b\in{\cal B}_{\delta}\}$ and $W^k_{\delta}(b,A) = V^k(b,A)$ for all $b\in{\cal B}_{\delta}$. We can consider the standard dynamic programing with the discretized state space  as a special case of the update rules in \eqref{eq:nblm3}-\eqref{eq:avf}.  Then, the performance achieved with the approximate value function can be characterized as follows.

\begin{proposition}\label{pp:pfbd}
The approximate value function obtained by recursively solving \eqref{eq:nblm3}-\eqref{eq:avf} is no less  than the discrete value function obtained by the standard dynamic programming method with the state space  ${\cal B}_{\delta}\times \{0,1\}$ where $\delta$ is the approximate precision.
\end{proposition}
\begin{IEEEproof}
Given the discrete state space ${\cal B}_{\delta}\times \{0,1\}$, since $W_\delta^{(i)}(B,A) = V^{(i)}(B,A)$ for any $B\times A\in{\cal B}_{\delta}\times \{0,1\}$, the standard dynamic programming follows the same update rule in \eqref{eq:nblm3}-\eqref{eq:avf} but with a discrete feasible power allocation set for the optimization problem in \eqref{eq:nblm2}, which is a subset of $D_1(b)$. 
\end{IEEEproof}

Moreover, in the standard discrete dynamic programming, we discretize all continuous variables, i.e., $b_k,h_k,e_k,p_k$, and then perform the dynamic programming with an exhaustive search on $p_k$ for all possible combinations of $(b_k,h_k)$; while with the proposed approximate value function, we only discretize the battery state $b_k$ and then obtain the approximate value function for each discretized $b_k$ in closed-form.

\subsection{Concavity of Approximate Value Function}
In \eqref{eq:nblm3}-\eqref{eq:avf}, we note that the approximate value function is based on the solution to an optimization problem \eqref{eq:nblm2}. To facilitate solving  \eqref{eq:nblm2}, in this subsection, we will show that the approximate value function $W_{\delta}^k(b,A)$ given in \eqref{eq:avf} is concave for $0\leq b\leq b_{\max}$ given $A \in \{0,1\}$. Then \eqref{eq:nblm2} is a  convex optimization problem given $h$ and $b$.

\begin{figure}
\centering
\includegraphics[width=0.6\textwidth]{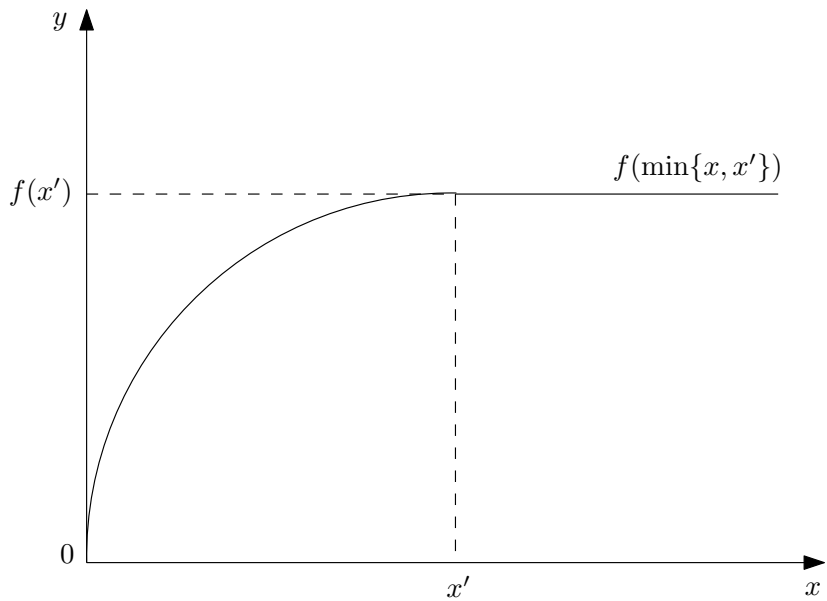}
\caption{Illustration of Lemma \ref{lm:init}. }
\label{fg:lm1}
\end{figure}

First, we introduce the following lemma, which can be easily shown and illustrated in Fig. \ref{fg:lm1}.
\begin{lemma}\label{lm:init}
If a function $f(x)\in\mathbb{R}\ (x\in\cal X\subseteq \mathbb{R})$ is non-decreasing, for any $x'\in{\cal X}$, $f(\min\{x,x'\})$ is also non-decreasing. Further, if the non-decreasing function $f(x)$ is concave, then $f(\min\{x,x'\})$ is concave for $x\in {\cal X}\cup [x',\infty)$.
\end{lemma}

We have the following  non-decreasing property  of $W_{\delta}^{k}(b,A)$.
\begin{proposition}\label{pp:nd}
For any $k \in\{1,2,\ldots, K-1\}$, if the approximate value function $W_{\delta}^{k+1}(b,A)$ is non-decreasing with respect to $b\in[0,b_{\max}]$ given $A\in\{0,1\}$, so is $W_\delta^k(b,A)$.
\end{proposition}
\begin{IEEEproof}
If $W_{\delta}^{k+1}(b,A)$ is non-decreasing with respect to $b\in[0,b_{\max}]$ for $A\in\{0,1\}$, by Lemma \ref{lm:init}, we have that $W_{\delta}^{k+1}(\min\{b_{\max},b\},A)$ is also non-decreasing with respect to $b\in[0,+\infty)$. Then, we have that $ U^{k}(b,p,A)$, which is a linear combination of the terms of the form $W_{\delta}^{k+1}(\min\{b_{\max},b+e_k -p_kT_c\},A)$, is also non-decreasing  with respect to $b\in[0,b_{\max}]$, given $p$ and $A$.

Given any battery level $b\in[0,b_{\max})$, channel fading $h$, the power $p_0$ such that $p_0\in{\cal D}_A(b)$, and $\epsilon>0$ such that $b + \epsilon \leq b_{\max}$, we have
\begin{equation}
p_0\in{\cal D}_{A}(b+\epsilon)\ ,
\end{equation}
and
\begin{align}
\log(1+p_0h) + \gamma  U^{k}(b,p_0,A) &\leq \log(1+p_0h) + \gamma  U^{k}(b + \epsilon,p_0,A)\\
& \leq \max_{p\in{\cal D}_{A}(b+\epsilon)} \big\{ \log(1+ph)  + \gamma  U^{k}(b+\epsilon,p,A)\big\}\ .
\end{align}

Since $V^k(b,A)$ is a non-negative linear combination of the terms of the form $\max_{p\in{\cal D}_A(b)} \big\{ \log(1+ph) +  U^{k}(b,p,A)\big\}$, $V^k(b,A)$ is non-decreasing with respect to $b\in[0,b_{\max}]$. Then, by \eqref{eq:avf}, we have that $W_{\delta}^k(b,A)$ is also non-decreasing with respect to $b\in[0,b_{\max}]$.
\end{IEEEproof}

The next result is on the concavity of $W_{\delta}^{k}(b,A)$.

\begin{proposition}\label{pp:concave}
For any $k \in \{ 1,2,\ldots,K\}$, if the approximate value function $W_{\delta}^{k+1}(b,A)$  is non-decreasing and concave with respect to $b\in[0,b_{\max}]$ given $A\in\{0,1\}$, so is $W_{\delta}^{k}(b,A)$.
\end{proposition}
\begin{IEEEproof}
Since  $W_{\delta}^{k+1}(b,A)$  is non-decreasing and concave with respect to $b\in[0,b_{\max}]$ given $A\in\{0,1\}$, by Lemma \ref{lm:init}, we have $W_{\delta}^{k+1}(\min\{b_{\max},b \},A)$  is non-decreasing and concave with respect to $b\geq 0$ given $A\in\{0,1\}$. Since $b + e  - pT_c$ is a linear combination of $b$ and $p$, then $W_{\delta}^{k+1}(\min\{b_{\max},b + e  - pT_c\},A)$ is jointly concave with respect to $b$ and $p$. Moreover, it follows that $U^k(b,p,A)$ is also jointly concave with respect to  $b$ and $p$ given $A\in\{0,1\}$\cite{CO}.

Since the feasible domain ${\cal D}_A(b)$ is different under $A=0$ and $A=1$. We consider the two cases separately.

When $A=0$, since ${\cal D}_0=0$, $v^k(b,0)$ can be written as
\begin{equation}
V^{k}(b,0) =  \mathbb{E}_{h_k}\Big[ \gamma U^k(b,0,0)\Big] \ .
\end{equation}
Since $U^k(b,p,A)$  is concave with respect to $b\in[0,b_{\max}]$ given $p$ and $A\in\{0,1\}$, so is $V^{k}(b,0)$ \cite{CO}. Then, by \eqref{eq:avf}, $W_{\delta}^k(b,0)$ is non-decreasing with respect to $b\in[0,b_{\max}]$.

When $A=1$, the feasible domain of the objective function in \eqref{eq:blm2}  is given by  ${\cal C} \triangleq \{(b,p)\;:\; 0\leq b\leq b_{\max},0\leq p\leq \min\{b/T_c,  p_{\max}\}\}$. It can be verified that $\cal C$ is a convex set. Then, for any $(b_1,p_1),(b_2,p_2)\in{\cal C}$, their convex combination $(\theta b_1 + \bar{\theta}b_2,\theta p_1 + \bar{\theta}p_2)\in{\cal C}$, where $\theta\in[0,1]$ and $\bar{\theta}\triangleq1-\theta$.

Moreover,  since ${\cal D}_1(b_1),{\cal D}_1(b_2)$ are non-empty, we can denote
\begin{align}
p_1 = \arg \max_{p\in{\cal D}_1(b_1)}\Big\{ \log(1+ph)  +  \gamma  U^{k}(b_1,p,1)\Big\}\ ,\label{eq:11}
\end{align}
and
\begin{align}
p_2 = \arg \max_{p\in{\cal D}_1(b_2)}\Big\{ \log(1+ph)  +  \gamma  U^{k}(b_2,p,1)\Big\}\ .\label{eq:12}
\end{align}
Then
\begin{align}
&\max_{p\in{\cal D}_1(\theta b_1 + \bar{\theta}b_2)}\Big\{ \log(1+ph)  +  \gamma  U^{k+1}(\theta b_1 + \bar{\theta}b_2,p,1)\Big\}\nonumber\\
&\leq \log(1+(\theta p_1 + \bar{\theta}p_2)h) +\gamma  U^{k+1}(\theta b_1 + \bar{\theta}b_2,\theta p_1 + \bar{\theta}p_2,1)\nonumber\\
&\leq \theta \log(1+p_1h)  + \bar{\theta} \log(1+p_2h)  + \theta \gamma  U^{k+1}(b_1, p_1,1) + \bar{\theta} \gamma  U^{k+1}(b_2, p_2,1)\label{eq:1}\\
&= \theta\big( \log(1+p_1h)  + \gamma  U^{k+1}(b_1, p_1,1)\big) + \bar{\theta}\big( \log(1+p_2h)  + \gamma   U^{k+1}(b_2, p_2,1)\big)\nonumber\\
&= \theta\max_{p\in{\cal D}_1(b_1)}\Big\{ \log(1+ph)  +  \gamma U^{k+1}(b_1,p,1)\Big\} + \bar{\theta}\max_{p\in{\cal D}_2(b_2)}\Big\{ \log(1+ph)  +  \gamma U^{k+1}(b_2,p,1)\Big\}\label{eq:2}\ ,
\end{align}
where \eqref{eq:1} follows from the joint  concavity, and \eqref{eq:2} follows from the definitions in \eqref{eq:11} and \eqref{eq:12}.

Therefore, we have that $\max_{p\in{\cal D}_1(b)}\big\{ \log(1+ph)  +  \gamma  U^{k+1}(b,p,1)\big\}$ is concave with respect to $b\in[0,b_{\max}]$. By \eqref{eq:nblm2} and  \eqref{eq:avf}, we further have $W_\delta^k(b,1)$ is concave with respect to $b\in[0,b_{\max}]$ \cite{CO}.
\end{IEEEproof}

From Propositions \ref{pp:nd} and \ref{pp:concave}, we have that if $W_\delta^{k+1}(b,A)$ is non-decreasing and concave so is $W_\delta^k(b,A)$ for any $k\in\{1,2,\ldots,K-1\}$.  Since $\log(1+ph)$ is non-decreasing and concave with respect to  $b\in[0,b_{\max}]$, it is easily verified by \eqref{eq:blm2} that $W_{\delta}^{K}(b,A)=V^K(b,A)=v^K(b,A)$ is also non-decreasing and concave with respect to  $b\in[0,b_{\max}]$ given $A$. By induction, we obtain the following theorem.
\begin{theorem}\label{thm:convex}
For $k=1,2,\ldots, K$, the approximate value function $W_\delta^k(b,A)$ is non-decreasing and concave with respect to $b\in[0,b_{\max}]$ given $A\in\{0,1\}$. Further, the problem in \eqref{eq:nblm2} is a convex optimization problem given $b\in[0,b_{\max}]$ and $A\in\{0,1\}$.
\end{theorem}

\begin{figure}[!h]
\centering
\includegraphics[width=0.75\textwidth]{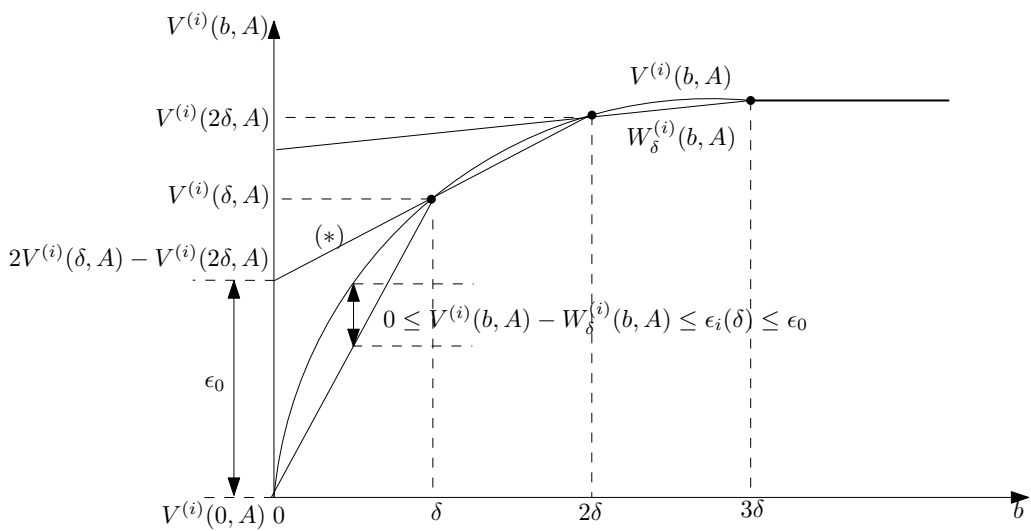}
\caption{The piecewise linear approximation of the value function and the approximation error bound. }
\label{fg:bd}
\end{figure}

Since both $V^{(i)} (b,A)$ and  $W_{\delta}^{(i)}(b,A)$ are concave and non-decreasing, where $i=K-k+1$ is the iteration number, we can further bound the approximation error  $ \epsilon_i(\delta)$ in \eqref{eq:err}  as follows.
\begin{proposition}\label{pp:err}
For any iteration $i$, given $A$, we have
\begin{equation}\label{eq:errbound}
0\leq \epsilon_i(\delta) \leq 2V^{(i)}(\delta,A)  - V^{(i)}(2\delta,A)- V^{(i)}(0,A)\ .
\end{equation}
\end{proposition}
\begin{IEEEproof}
By Theorem \ref{thm:convex}, $V^{(i)}(b,A)$ is non-decreasing and concave with respect to $b$ given $A$. As illustrated in Fig.~\ref{fg:bd}, for $b\in[0,\delta]$, the value of $V^{(i)}(b,A)$ is smaller than the value on line (*) but larger than $W_\delta^{(i)}(b,A)$, and therefore the distance between the value on line (*) and $W_\delta^{(i)}(b,A)$ can also be considered as an upper bound on the approximation  error, i.e., $V^{(i)}(b,A)-W_\delta^{(i)}(b,A)$ for $b\in[0,\delta]$. According to the second-order derivative property of the concave function, we have that
\begin{align}
&V^{(i)}((n+1)\delta,A)- V^{(i)}(n\delta,A)  - (V^{(i)}((n+2)\delta,A)  - V^{(i)}((n+1)\delta,A))\nonumber \\
\geq& V^{(i)}((n+2)\delta,A)  - V^{(i)}((n+1)\delta,A) - (V^{(i)}((n+3)\delta,A)  - V^{(i)}((n+2)\delta,A))
\end{align}
for all $n\geq 0$. Then, we further have that $0\leq \epsilon_i(\delta) \leq \max\{2V^{(i)}(\delta,A)  - V^{(i)}(2\delta,A)- V^{(i)}(0,A),2V^{(i)}(2\delta,A)  - V^{(i)}(3\delta,A)- V^{(i)}(\delta,A),\cdots\}=2V^{(i)}(\delta,A)  - V^{(i)}(2\delta,A)- V^{(i)}(0,A)$, where $\epsilon_i(\delta)=||V^{(i)}(b,A)-W_\delta^{(i)}(b,A)||_\infty$.
\end{IEEEproof}

\section{Power Allocation with Prefect Energy Prediction}
Note that  in \eqref{eq:nblm2}, we need to solve the following optimization  problem for a given $B\in{\cal B}_{\delta}$ and $A\in\{0,1\}$:
\begin{equation}\label{eq:sproblem}
p^*(h)=\arg\max_{p(h)\in{\cal D}_A(B)} \big\{ \log(1+p(h)h) + \gamma  U^{k}(B,p(h),A) \big\},\  h\geq 0\ .
\end{equation}

When $A=0$, $p^*(h)=0$. On the other hand, when $A=1$, we will obtain the optimal solution  $p^*(h)$ in closed-form.

Since the approximate value function $W_{\delta}^{k+1}(b,A)$ in \eqref{eq:avf} is a piecewise linear function of $b$ given $A$, it follows that $U^k(B,p,1)$ in \eqref{eq:nblm3} is also a piecewise linear function with respect to $p$ given $B$, which is differentiable everywhere except at ${\cal J}\triangleq\{p\;|\;p=(B+e_k-B_0)/T_c,B_0\in{\cal B}_{\delta}\}$. By Theorem \ref{thm:convex} and Lemma \ref{lm:init}, $U^k(B,p,1)$ is also concave and non-decreasing with respect to $p$.

\begin{figure}[!htp]
\centering
\includegraphics[width=0.6\textwidth]{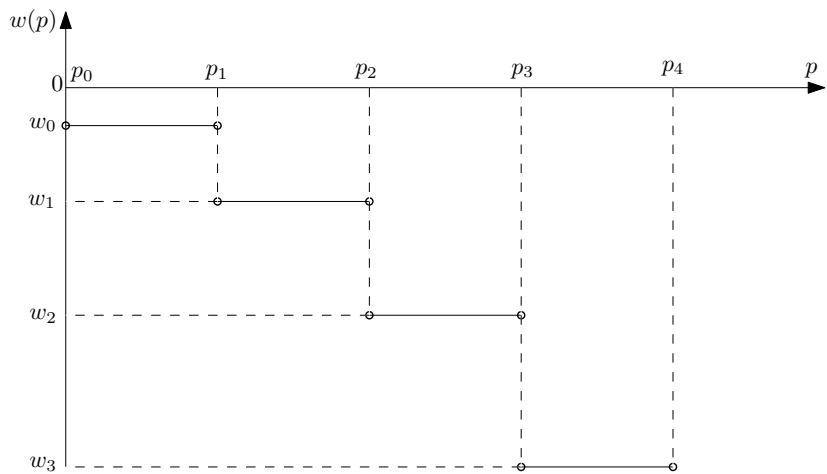}
\caption{The derivative of $U^k(B, p, 1)$ with respect to $p$.}
\label{fg:staircase}
\end{figure}

Since $U^k(B,p,1)$ is  a piecewise linear function, we denote ${\cal I}\triangleq \{p_0,p_1,\ldots,p_N\}$ as the set of the non-differentiable points, where $p_0=0$, $p_N=\min\{p_{\max},B/T_c\}$, and $p_i,(0<i<N)$ is the $i$-th smallest element in ${\cal J} \cap {\cal D}_1(B) \setminus\{p_0,p_N\} $. Also, we denote ${\cal W}=\{w_1,w_2,\ldots, w_N\}$ as the set of the corresponding slopes, where $w_i$ is the slope of the segment $[p_{i-1},p_i]$, given by
\begin{align}
w_i \triangleq
-\frac{\gamma T_c}{\delta}\mathbb{E}_{A\;|\;1}\Big\{ & V^{k+1}(\left\lceil \min\{b_{\max},B + e_k - p_iT_c\}/ \delta\right\rceil \delta,A)\nonumber\\
&-V^{k+1}(\left\lfloor\min\{b_{\max},B + e_k - p_iT_c\}/ \delta\right\rfloor \delta,A)\Big\}\ ,
\end{align}
which is derived from \eqref{eq:nblm3} and  \eqref{eq:avf}. Hence, the derivative of $U^k(B,p,1)$ for $p\in{\cal D}_1(B) \setminus{\cal I}$ is
\begin{equation}\label{eq:ud}
w(p)=w_i, \textrm{ if }p\in(p_{i-1},p_i)\ .
\end{equation}
Since $U^k(b,p,A)$ is concave and non-decreasing with respect to $p$, we have $0\geq w_0 > w_1 > \ldots > w_N$.  Fig. \ref{fg:staircase} is a sketch of the stair-case function $w(p)$.

In this section we first obtain the closed-form solution to \eqref{eq:sproblem}, and then use it to obtain the optimal power allocation for both finite- and infinite-horizon cases.

\subsection{The Optimal Solution to \eqref{eq:sproblem}}
In this subsection, for simplicity, we drop the superscript $k$ and denote the objective function in \eqref{eq:sproblem} as
\begin{equation}
g_h(p) \triangleq \log(1+ph) + \gamma  U(B,p,1),\  p\in {\cal D}_1(B)\ .
\end{equation}
We note that $g_h(p)$ is differentiable for $p\in {\cal D}_1(B)\setminus{\cal I}$ with
\begin{equation}\label{eq:dev}
g_h'(p) = \frac{1}{1/h+p} + w(p)\ .
\end{equation}
On the other hand, at the non-differentiable points in ${\cal I}$, the right-derivative and the left-derivative of $g_h(p)$ can be written as
\begin{equation}\label{eq:dev-}
{{g}_h'}(p^+) \triangleq \frac{1}{1/h+p} + w(p^+)\ ,
\end{equation}
and
\begin{equation}\label{eq:dev+}
{{g}_h'}(p^-) \triangleq  \frac{1}{1/h+p}+ w(p^-)\ ,
\end{equation}
respectively.
\begin{theorem}\label{thm:optsol}
The optimal solution to \eqref{eq:sproblem} is given by
\begin{equation}\label{eq:sum}
p^*(h)=\left\{\begin{array}{ll}
-\frac{1}{w_i} - \frac{1}{h}& \frac{1}{h}\in [-\frac{1}{w_{i}}-p_{i},-\frac{1}{w_{i}}-p_{i-1}]\cap [0,+\infty), i=1,2,\ldots,N-1\\
p_i& \frac{1}{h}\in(-\frac{1}{w_{i+1}}-p_{i},-\frac{1}{w_i}-p_i)\cap[0,+\infty), i=1,2,\ldots,N-1\\
0&\frac{1}{h} \in (-\frac{1}{w_1}-p_0,\infty)\\
p_N & \frac{1}{h} \in [0, -\frac{1}{w_N}-p_N) \\
\end{array}\right.\ ,
\end{equation}
where $p_0 = 0$ and $p_N=\min\{p_{\max},B/T_c\}$.
\end{theorem}

\begin{figure}
\centering
\includegraphics[width=0.6\textwidth]{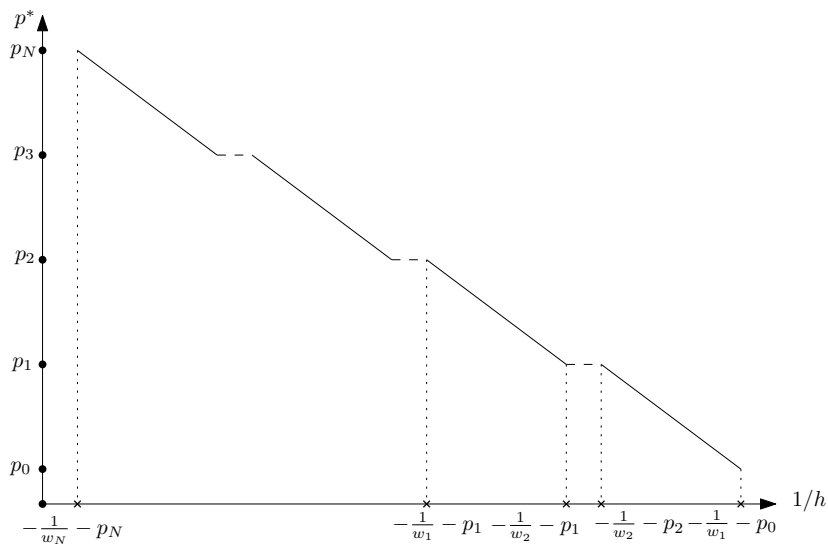}
\caption{The optimal solution  $p^*(h)$.}
\label{fg:wf}
\end{figure}

In Fig. \ref{fg:wf} we give a sketch of $p^*(h)$. To prove Theorem \ref{thm:optsol}, we first give the necessary and sufficient conditions for the optimal solution $p^*$ as follows \cite{CO}.
\begin{lemma}\label{lm:opt}
$p^*$ is the optimal solution to \eqref{eq:sproblem} given $h$, if and only if,
\begin{enumerate}
\item ${{g}'_h}({p^*}^+) \leq 0 \leq{{g}'_h}({p^*}^-)$, when  ${{g}'_h}(0^+) > 0$ and  ${{g}'_h}(\min\{B/T_c,p_{\max}\}^-) < 0$;
\item $p^* = \min\{B/T_c,p_{\max}\}$, when ${{g}'_h}(\min\{B/T_c,p_{\max}\}^-) \geq 0$;
\item $p^* = 0$, when ${{g}'_h}(0^+) \leq 0$.
\end{enumerate}
\end{lemma}
Note that, Condition 1 corresponds to  the case that  $p^*$ is in the interior of ${\cal D}_1(B)$. In this case, the left-derivative and the right-derivative should have opposite signs or be both zero at $p^*$ so that the increasing and decreasing of $p$ both lead to the decreasing of the objective function. Condition 2 and Condition 3 correspond to the cases that $p^*$ is on each side of the boundary of ${\cal D}_1(B)$, where the objective function is non-decreasing and non-increasing for all $p\in{\cal D}_1(B)$, respectively.


The following proposition gives a sufficient condition for the optimality of $p^*(h)$ given $B$.
\begin{proposition}\label{pp:in}
Given any $B\in{\cal B}_\delta$, for $h\geq 0$, if the energy schedule $p^*(h)\in int{\cal D}_{1}(B)$ satisfies
\begin{equation}\label{eq:in}
p^*(h)=\left\{\begin{array}{ll}
-\frac{1}{w(p^*(h))} -\frac{1}{h},& \textrm{ when } p^*(h)\in int{\cal D}_1(B) \setminus{\cal I},\\
-\frac{1}{w(p^*(h)^-)} -\frac{1}{h}  \textrm{ or } -\frac{1}{w(p^*(h)^+)} -\frac{1}{h} ,& \textrm{ when } p^*(h)\in{\cal I},\\
\end{array}\right.
\end{equation}
then $p^*(h)$ is the optimal solution to \eqref{eq:sproblem}.
\end{proposition}
\begin{IEEEproof}
Substituting \eqref{eq:in} into \eqref{eq:dev-}-\eqref{eq:dev+}, we have ${{g}'_h}(p^*(h)^+) = 0$ or ${{g}'}(p^*(h)^-)=0$ when  $p^*(h)\in{\cal I}$, and  ${{g}'_h}(p^*(h)^+) = {g}'(p^*(h)^-)=0$ when $p^*(h)\in int{\cal D}_1(B) \setminus{\cal I}$. Since  ${{g}'_h}(p^*(h)^+) \leq {{g}'_h}(p^*(h)^-)$, we  have ${{g}'_h}(p^*(h)^+) \leq 0 \leq {{g}'_h}(p^*(h)^-)$. Moreover, since $g_h(p)$ is concave, we  have $0\leq g'_{h}(p^*(h)^-)< g'_{h}(0^-)$ and $g'_{h}(\min\{p_{\max},B/T_c\}^-)< g'_{h}(p^*(h)^+)\leq 0$. By Lemma \ref{lm:opt} (Condition 1), we conclude the optimality.
\end{IEEEproof}

Then it is easy to verify that for  $\frac{1}{h}\in [-\frac{1}{w_{i}}-p_{i},-\frac{1}{w_{i}}-p_{i-1}]\cap [0,+\infty), i=1,2,\ldots,N-1$, the solution given by \eqref{eq:sum} satisfies the optimality condition in Proposition \ref{pp:in}.

For $\frac{1}{h}\in(-\frac{1}{w_{i+1}}-p_{i},-\frac{1}{w_i}-p_i)\cap[0,+\infty), i=1,2,\ldots,N-1$, we use the next proposition to prove the optimality of \eqref{eq:sum}.
\begin{proposition}\label{pp:fg}
For any non-differentiable point $p_i\in{\cal I} \setminus \{p_0,p_N\}$, $p_i$ is the optimal solution to \eqref{eq:sproblem} for any $\frac{1}{h}\in(-\frac{1}{w_{i+1}}-p_{i},-\frac{1}{w_i}-p_i)\cap[0,+\infty)$.
\end{proposition}
\begin{IEEEproof}
From \eqref{eq:dev-}-\eqref{eq:dev+}, ${g'_h}(p_i^+)$ and ${g'_h}(p_i^-)$ are functions of $\frac{1}{h}$ for a given $p_i$. If $(-\frac{1}{w_{i+1}}-p_{i},-\frac{1}{w_i}-p_i)\cap[0,+\infty)$ is not empty, it is easy to verify that  $0 =  {g'_h}({p_i}^-) > {g'_h}({p_i}^+) $ when $\frac{1}{h}=-\frac{1}{w_i}-p_i$, and $ {g'_h}(p_i^-) > {g'_h}(p_i^+)$ and ${g'_h}_+(p_i) \leq 0$ when $\frac{1}{h}=-\frac{1}{w_{i+1}}-p_i$. Since given $p_i$, ${g'_h}(p_i^-)$ and ${g'_h}(p_i^+)$ increase as $\frac{1}{h}$ decreases, then decreasing $\frac{1}{h}$ from $-\frac{1}{w_{i}}-p_{i}$ to $\max\{0,-\frac{1}{w_{i+1}}-p_{i}$\},  we have ${g'_h}(p_i^-)\geq 0\geq {g'_h}(p_i^+)$ for all $\frac{1}{h}\in(-\frac{1}{w_{i+1}}-p_{i},-\frac{1}{w_i}-p_i)\cap[0,+\infty)$. By Lemma \ref{lm:opt}, the proposition follows.
\end{IEEEproof}

Propositions \ref{pp:in} and \ref{pp:fg} obtain the optimal solution for $\frac{1}{h} \in[-\frac{1}{w_N}-p_N,-\frac{1}{w_1}]\cap [0,\infty)$. For other $h\geq 0$, using Conditions 2 and 3 in Lemma \ref{lm:opt}, we can prove the optimality of \eqref{eq:sum} as follows.
\begin{proposition}\label{pp:bl}
\begin{enumerate}
\item For any $h$ such that $\frac{1}{h} \geq -\frac{1}{w_1}$, the optimal solution is $p^*(h)=0$;
\item For any $h$ such that $0\leq \frac{1}{h}\leq -\frac{1}{w_N}-p_N $, the optimal solution $p^*(h)=p_N$.
\end{enumerate}
\end{proposition}
\begin{IEEEproof}
Note that, since $U(B,p,1)$ is non-increasing with respect to $p$, we have $-\frac{1}{w_1}\geq 0$. When $\frac{1}{h}=-\frac{1}{w_1}$, it is easy to verify that ${g'_h}(0^+) = 0$. Since ${g'_h}(0^+)$ is also a function of $\frac{1}{h}$ which decreases as $\frac{1}{h}$ increases, we have for any $\frac{1}{h}\geq -\frac{1}{w_1}$, ${g'_h}(0^+)\leq 0$. By Condition 3 in Lemma \ref{lm:opt}, we must have $p^*(h)=0$ for any $h$ such that $\frac{1}{h} \geq -\frac{1}{w_1}\geq 0$. Similarly, we may also verify that for any $\frac{1}{h}\leq -\frac{1}{w_N}-p_N$, ${g'_h}(p_N^-)\geq 0$. By Condition 2 in Lemma \ref{lm:opt}, we must have $p^*(h)=p_N$ for any $h$ such that $0 \leq \frac{1}{h} \leq  -\frac{1}{w_N} - p_N$.
\end{IEEEproof}

Note that, given $B\in{\cal B}_{\delta}$, $p^*(h)$ is a piecewise function in closed-form. Then, $V^k(B,A)$ can be efficiently evaluated as
\begin{equation}\label{eq:up1}
V^k(B,1)= \mathbb{E}_{h_k}\Big\{\log(1+p^*(h_k)h_k) + \gamma  U^{k}(B,p^*(h_k),1)\Big\}\ ,
\end{equation}
and
\begin{equation}\label{eq:up0}
V^k(B,0) = \mathbb{E}_{h_k}\Big\{\log(1) + \gamma U^k(B,0,0)\Big\}=U^k(B,0,0)\ ,
\end{equation}
where
\begin{equation}\label{eq:slinear}
U^{k}(B,p,0) = \sum_{A'=0,1}\textrm{Pr}(A_{k+1}=A'\;|\;A_{k}=0)\big[W_{\delta}^{k+1}(\min\{b_{\max},B + e_k - pT_c\},A') \big]\ .
\end{equation}
\subsection{Calculating the Approximate Value Function}
In order to obtain the power allocation, we need to compute the approximate value function given by \eqref{eq:nblm3}-\eqref{eq:avf} for $k=1,2,\ldots, K$ ($K=\infty$ for infinite-horizon case). Then, when the observation is available, we solve the problem given in \eqref{eq:mdpeq}.

\subsubsection{Finite $K$}
We first consider the finite-horizon case where $K$ is finite, we assume that the distributions of channel fading are independent across slots but not necessarily identical.

The power allocation consists of two  phases. In the first phase, we recursively compute the approximate value function from $k=K$ to $k=1$, following \eqref{eq:nblm3}-\eqref{eq:avf}. Specifically, in the $i$-th iteration, we obtain $W_\delta^{(i) }(b, A)$ for slot $k = K-i+1$ as follows. Based on $W_{\delta}^{(i-1)}(b,A)$ obtained in the previous iteration (or the initial function for the first iteration), for each $B\in{\cal B}_{\delta}$ and $A=\{0,1\}$, we obtain the piecewise linear function $U^{(i)}(B,p,A)$ by specifying the sets ${\cal I}$ and ${\cal W}$. Then, we use \eqref{eq:sum} to obtain $p^*(h)$ and use \eqref{eq:up1}-\eqref{eq:up0} to update $V^{(i)}(B,A)$ for all $B\in{\cal B}_{\delta}$ and $A=\{0,1\}$. With the set $\{V^{(i)}(B,A)\;|\;B\in{\cal B}_{\delta},A=\{0,1\}\}$, the approximate value function $W_{\delta}^{(i)}(b,A)$ can be obtained using \eqref{eq:avf} and we store the closed-form $W_{\delta}^{(i)}(b,A)$ in a look-up table. Note that the above  first phase should be completed before the first slot.

The second phase is performed at the beginning of each slot, once the observation becomes available. This phase is to solve the problem given in \eqref{eq:mdpeq} using \eqref{eq:sum}. Specifically, at the beginning of slot $k$, the transmitter observes the system state, i.e., the channel access state $A$, the channel gain $h$, and the current battery state $b$. When $A=0$, the transmitter keeps silent. Otherwise, the transmitter retrieves the approximate value function $W_{\delta}^{k+1}(b,A)$ (i.e., $W_{\delta}^{(K-i-1)}(b,A)$)  from the look-up table and then calculate the power allocation using \eqref{eq:sum}.

The entire computational procedure for the finite-horizon case is summarized in Algorithm 1. \\
\begin{minipage}[h]{6.5 in}
\rule{\linewidth}{0.3mm}\vspace{-.1in}
{\bf {\footnotesize Algorithm 1 - Finite-Horizon Power Allocation }}\vspace{-.2in}\\
\rule{\linewidth}{0.2mm}
{ {\small
\begin{tabular}{ll}
	\;1:&  Inputs\\
	\;& Distributions of $\boldsymbol{H}$, $\boldsymbol{A}$; value of $e_k$ for $k=1,2,\ldots,K$\\
    \;& The approximation precision  $\delta>0$ and the discount factor $\gamma\in[0,1]$.\\
    \;2:& Phase-I: Compute the approximate  value function update (offline calculation)\\
    	\;& {\bf FOR $k=K$ TO $1$}\\
    	\;(*)&\quad Calculate $V^{k}(B,h,A)$ for $B\in{\cal B}_{\delta},A\in\{0,1\}$ using  \eqref{eq:sum} and \eqref{eq:up1}-\eqref{eq:up0}\\
      \;&\quad Compute $W_{\delta}^{k}(b,A)$ from $V^{k}(B,A)$ using \eqref{eq:avf}  and store it\\
      	\;&{\bf ENDFOR}\\
    \;3;& Phase-II: Power Allocation (online calculation)\\
    \;&{\bf FOR $k=1$ TO $K$}\\
    \;& \quad Get the observations $\Gamma_k=(b_k,h_k,A_k)$\\
    \;& \quad Retrieve $W^{k+1}_\delta(b,A)$ and calculate $U^k(b_k,p,A_k)$ using \eqref{eq:nblm3} \\
    \;&\quad Calculate $p^*(h_k)$ using \eqref{eq:sum}\\
	\;&{\bf ENDFOR}\\
\end{tabular}}}\\
\rule{\linewidth}{0.3mm}
\end{minipage}\vspace{.2 in}\\

\begin{remark}
If the observations can be predicted in a scheduling period $K$, i.e., $\boldsymbol{H}$, $\boldsymbol{E}$, and $\boldsymbol{A}$ are known in advance, we can rewrite \eqref{eq:problem} as follows
\begin{equation}\label{eq:dtmproblem}
{\cal P}^* = \arg \max_{p_k,k=1,2,\ldots,K}\Big\{\sum_{k=1}^K A_k\log(1+p_kh_k)\Big\}\ ,
\end{equation}
subject to the constraints in \eqref{eq:powerc}, \eqref{geq:battery}, and \eqref{eq:batteryc} for $k = 1,2,\ldots,K$.

We note that in the above case all the observations are non-causally known in advance and the problem in \eqref{eq:dtmproblem} is a convex optimization problem. Instead of the generic convex solver, there is also an efficient dynamic water-filling algorithm proposed in \cite{MyPaper}, for solving \eqref{eq:dtmproblem} optimally. Moreover, since \eqref{eq:dtmproblem} is a special case of the stochastic case, Algorithm 1 is also applicable and would approach the optimal performance as the dynamic water-filling algorithm when $\delta \to 0$. Specifically, the use of Algorithm 1 or the dynamic water-filling algorithm strikes a balance between the performance and the computational complexity.
\end{remark}

\subsubsection{Infinite $K$}

In the infinite-horizon case, although $K$ is infinite, the number of the iterations in the first phase  is not infinite since the approximate value function will converge. Moreover, since we have assumed that $e_k$ is static and $h_k$ is i.i.d., the converged approximate value function can be directly used in \eqref{eq:mdpeq} to obtain the power allocation with the observations in the second phase, for all slots.

We denote
\begin{equation}
{\cal T}_\delta\;:\; W_{\delta}(b,A)\rightarrow W_{\delta}(b,A)
\end{equation}
as the value function update operator in \eqref{eq:nblm3}-\eqref{eq:avf}: based on a given value function $W_{\delta}^{(i)}(b,A)$, it solves \eqref{eq:nblm2} to obtain $V^{(i)}(B,p,A)$ for $B\in{\cal B}_{\delta}$, and then generates the new approximate value function $W_{\delta}^{(i+1)}(b,A)$ by \eqref{eq:avf}. Then we can write
\begin{equation}\label{eq:vi}
W_{\delta}^{(i+1)}(b,A) \triangleq {\cal T}_{\delta}\Big[W_{\delta}^{(i)}(b,A)\Big], \ b\in[0,b_{\max}]\ .
\end{equation}
Note that ${\cal T}_0$ is the standard Bellman operator corresponding to  \eqref{eq:blm2}-\eqref{eq:blm3} without the value function approximation, i.e., $\delta=0$ \cite{MDP}.

Then the computational procedure  for the infinite-horizon case is summarized in Algorithm 2.

\begin{minipage}[h]{6.5 in}
\rule{\linewidth}{0.3mm}\vspace{-.1in}
{\bf {\footnotesize Algorithm 2 -  Infinite-Horizon Power Allocation}}\vspace{-.2in}\\
\rule{\linewidth}{0.2mm}
{ {\small
\begin{tabular}{ll}
	\;1:&  Inputs\\
	\;& Distributions of $h$, $A$; value of $e$ \\
    \;& The approximation precision $\delta>0$, the discount factor $\gamma\in(0,1)$, and the termination condition $\alpha$.\\
	\;2:& Phase-I Approximate value function update (offline calculation)\\
    \;& $i \leftarrow 0$\\
	\;&\textbf{\bf REPEAT}\\
		\;(*)&\quad $W^{(i+1)}_{\delta}(b,A) = {\cal T}_{\delta}\Big[W^{(i)}_{\delta}(b,A)\Big]$\\
        \;&\quad $i \leftarrow i + 1$\\
	\;&\textbf{\bf UNTIL} $||W_{\delta}^{(i)}(b,A)-W_{\delta}^{(i-1)}(b,A)||_{\infty} \leq \alpha$\\
    \;& $W_{\delta}^*(b,A) \leftarrow W_{\delta}^{(i)}(b,A)$\\
	\;3:&Phase-II Power Allocation (online calculation)\\
	\;& {\bf AT THE BEGINNING OF EACH SLOT}\\
    \;& \quad Get the observations $\Gamma=(b,h,A)$\\
    \;& \quad Retrieve $W_\delta^*(b,A)$ and calculate $U^*(b,p,A)$ using \eqref{eq:nblm3} \\
    \;&\quad Calculate $p^*(h)$ using \eqref{eq:sum}\\
\end{tabular}}}\\
\rule{\linewidth}{0.3mm}
\end{minipage}\vspace{.2 in}\\

To show the convergence of the approximate value function update, we first note that, by repeatedly performing  ${\cal T}_0$ on any initial value function, a converged value function can be obtained as follows  \cite{MDP}:
\begin{equation}
v^*(b,A)\triangleq  {\cal T}_{0}\cdot{\cal T}_{0}\cdot \ldots\Big[v^{(1)}(b,A)\Big]=  {\cal T}_{0}^{\infty}\Big[v^{(1)}(b,A)\Big].
\end{equation}
Extending the convergence of ${\cal T}_{0}$ to ${\cal T}_{\delta}$, we introduce the following lemma. The proof is given in Appendix A.
\begin{lemma}\label{lm:cond}
The operator ${\cal T}_{\delta}$ has the $\gamma$-contraction property, i.e., for any two functions $V_1(b,A)$ and $V_2(b,A)$, we have
\begin{equation}
||{\cal T}_{\delta}\Big[V_1(b,A)\Big] - {\cal T}_{\delta}\Big[V_2(b,A)\Big]||_{\infty} \leq \gamma||V_1(b,A)-V_2(b,A)||_{\infty}\ .
\end{equation}
\end{lemma}

It then follows that
\begin{align}
||T_{\delta}^{i+1}\Big[W_{\delta}^{(1)}(b,A)\Big] - {\cal T}_{\delta}^{i}\Big[W_{\delta}^{(1)}(b,A)\Big]||_{\infty}
&\leq \gamma^i ||{\cal T}_{\delta}\Big[W_{\delta}^{(1)}(b,A)\Big] - W_{\delta}^{(1)}(b,A)||_{\infty}\ ,
\end{align}
i.e., ${\cal T}_{\delta}^{i}\Big[W^{(1)}(b,A)\Big]$ converges as $i$ increases. Moreover, the error between the converged approximate value function and $v^*(b,A)$ is bounded as follows.
\begin{theorem}\label{thm:inferr}
If $||{\cal T}_\delta^i \left[W_{\delta}^{(1)}(b,A)\right] -  {\cal T}_\delta^{i-1} \left[W_{\delta}^{(1)}(b,A)\right]||_{\infty}\leq \alpha$, then the error between $v^*(b,A)$ and $W_{\delta}^{(i)}(b,A)$ is bounded by
\begin{align}
||W_{\delta}^{(i)}(b,A)-v^*(b,A)||_{\infty} &\leq  \frac{\gamma\alpha + ||2v^*(\delta,A) - v(0,A) - v^*(2\delta,A)||_{\infty} }{1-\gamma}\ .
\end{align}
\end{theorem}
\begin{IEEEproof}
The proof is provided in Appendix B.
\end{IEEEproof}

Note that,  Algorithms 1 and 2 have both the offline calculation part and the online calculation part. During offline calculation, we evaluate $V^k(B,A)$ for each $B\in{\cal B}_{\delta}$ in each iteration, i.e., solve ${\cal O}(B_{\max}/\delta)$ convex optimization problems in each iteration. Specifically, rather than using an exhaustive search for each combination of the discretized $(B,H)$ ($H$ is the discretized channel gain) as done by the standard discrete MDP method, the proposed algorithms use \eqref{eq:sum} to calculate $V^k(B,A)$ for each $B\in{\cal B}_{\delta}$ directly. Moreover, for the infinite case, by Lemma \ref{lm:cond}, the $\alpha$-converged approximate value function can be obtained within ${\cal O}(\log_\gamma\alpha)$ iterations. On the other hand, during online calculation, we retrieve $W^{k+1}_\delta(b,A)$ (or $W^*_\delta(b,A)$) from the look-up table and then use  \eqref{eq:sum} to compute the power allocation for the specific observation $(b_k,h_k,A_k)$.

Moreover, the proposed algorithms calculate the power allocation based on the continuous battery state and channel gain, and the obtained  power allocation is also continuous. Thus it provides higher precision for both offline calculation and online calculation than the conventional discrete MDP method, especially when the discretization step is large. Finally,  as shown in Section VI, a better performance can be achieved by the proposed algorithm with a lower computational complexity compared with the conventional discrete MDP method.


\section{Power Allocation with Imperfect Energy Prediction}
Although energy harvesting is usually predictable, there may exist a non-negligible prediction error in practice. In this section, we treat the case of imperfect energy harvesting  prediction where the prediction error is an i.i.d. random variable. We also consider a general payoff function $r(p,A)$, which is continuous, non-decreasing and concave with respect to $p$ given $A\in\{0,1\}$.

In this general model, we assume the energy harvesting process consisting of  a deterministic part  $e_k$ and a stochastic part  $\varepsilon_k$. The deterministic process $e_k$ in practice is obtained from the prediction using historic observations, e.g., by averaging  the historic measurement with the weather adjustment.


With the prediction error, the problem formulation is modified as follows:
\begin{equation}\label{geq:battery}
b_{k+1} = \min\big\{b_{\max}, b_k + e_k + \varepsilon_k - p_kT_c\big\} \ ,
\end{equation}
 and
\begin{equation}\label{geq:problem}
{\cal P}^* \triangleq \arg \max_{p_k(\cdot),k=1,2,\ldots,K}\Big\{\mathbb E_{\boldsymbol{H},\boldsymbol{E},\boldsymbol{A}} \Big[\sum_{k=1}^K \gamma^{k-1} r(p_k(\Gamma_k),h_k)\Big]\Big\}\ ,
\end{equation}
subject to the constraints in \eqref{eq:powerc},  \eqref{eq:batteryc}, and \eqref{geq:battery}, for $k = 1,2,\ldots,K$, where $\boldsymbol{E}\triangleq[\varepsilon_1,\varepsilon_2,\ldots,\varepsilon_K]$. Accordingly, since $\varepsilon_k$ is a random variable, the (approximate) value function update rules in \eqref{eq:blm3} and \eqref{eq:nblm3} are changed to
\begin{align}
 u^{k}(b_k,p_k,A_k) &\triangleq \mathbb{E}_{\varepsilon_k,A_{k+1}|A_k}\big[v^{k+1}(\min\{b_{\max},b_k + e_k + \varepsilon_k - p_kT_c\},A_{k+1}) \big]\label{geq:blm3}\ ,
\end{align}
and
\begin{align}
 U^{k}(b_k,p_k,A_k) &\triangleq \mathbb{E}_{\varepsilon_k,A_{k+1}|A_k}\big[W_{\delta}^{k+1}(\min\{b_{\max},b_k + e_k+ \varepsilon_k - p_kT_c\},A_{k+1}) \big]\ ,\label{geq:nblm3}
\end{align}
respectively.

Obviously, since $r(p,A)$ is continuous, non-decreasing and concave with respect to $p$ given $A$, and the expectation with respect to $\varepsilon_k$ in \eqref{geq:blm3} and \eqref{geq:nblm3} preserves the concavity and the non-decreasing properties, we can extend the analysis in Section III to the case with the general payoff function and imperfect energy prediction, obtaining the same concavity and non-decreasing properties.

However, note that, the optimal solution $p^*(h)$ in \eqref{eq:sum} is based on the facts that $ U^{k}(b,p,A)$ is a piecewise linear function and $r(p,h)=\log(1+ph)$, which are no longer valid with the general payoff function and/or imperfect energy prediction. Then, in Algorithms 1 and 2, the steps marked by (*), which aim to solve the problem in \eqref{eq:nblm2}, need to be modified accordingly. In particular, we now need to use some standard convex solver to numerically solve \eqref{eq:nblm2}.

\section{Simulation Results}
We use the payoff function $r(p,h)=\log(1+ph)$. We assume that the channel fading $h_k$ is an i.i.d. random variable following the Rayleigh distribution with the parameter $\sigma$. We first assume that the harvested energy can be perfectly predicted. For the transmitter, we set the maximum transmission power as $p_{\max}=6$ units per slot, the battery capacity as $b_{\max}=15$ units, and the initial battery level as $b_0=2$ units. Further, we set the probability of the channel access suspension as $q=\hat{q}=0.1$, the approximate precision $\delta$ of the approximating value function as $1$ and $0.1$, and the convergence error tolerance  for the infinite-horizon case as $\alpha=0.0001$.

We first evaluate the performance of the proposed algorithms. For comparison, we consider three simple power allocation methods, the {\em greedy policy}, the {\em balanced policy}, and the standard discrete MDP method. The greedy policy tries to allocate as much power as possible in each slot subject to the energy availability. On the other hand, the balanced policy tries to allocate a constant power in each slot, e.g., the mean value of the harvested energy.  Moreover, for the standard discrete MDP method, we discretize the battery level, the channel gain, and the transmission power with the same precision  factor $\delta$, and then perform the dynamic programming algorithm and the value iteration algorithm on the discrete state space  for the finite- and infinite-horizon cases, respectively.

\begin{figure}
\centering
\includegraphics[width=0.75\textwidth]{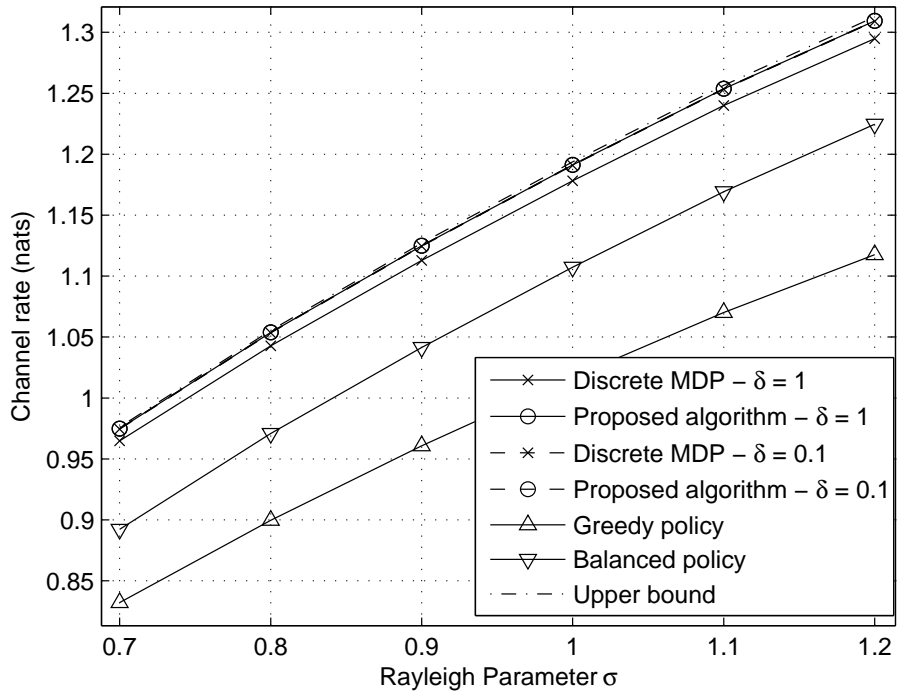}
\caption{Performance comparisons in the energy-constrained scenario for the finite-horizon case.}
\label{fg:fint_ec}
\end{figure}

\begin{figure}
\centering
\includegraphics[width=0.75\textwidth]{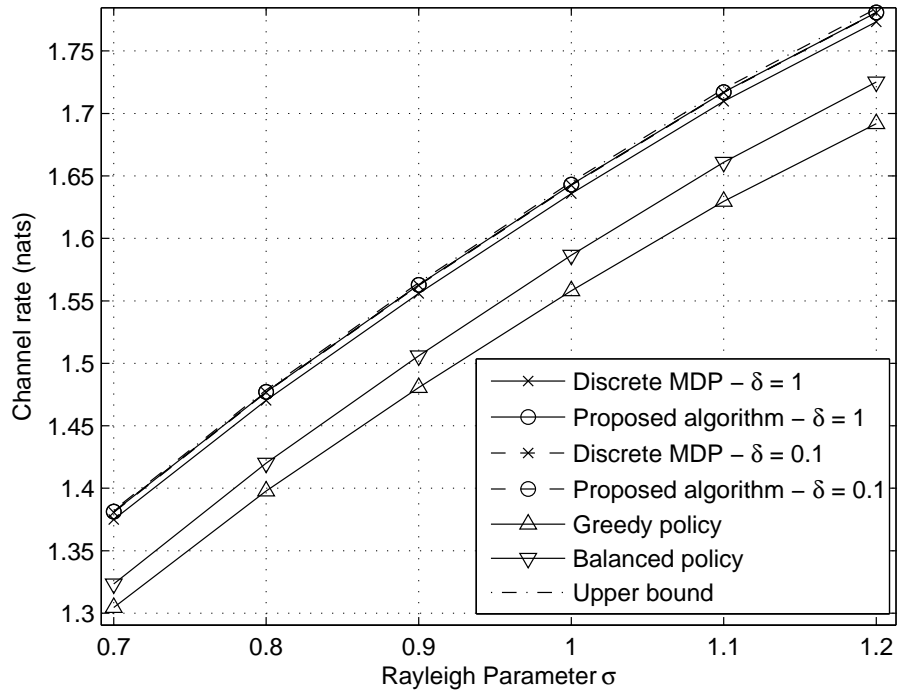}
\caption{Performance comparisons in the power-constrained scenario for the finite-horizon case.}
\label{fg:fint_pc}
\end{figure}

For the finite-horizon case, we set $K=30$, $\gamma=1$, and $\sigma=0.7,0.8,0.9,1.0,1.1,1.2$. We randomly generate the prediction value $e_k$ following a positive truncated-Gaussian distribution with the variance of $2$. We consider two typical scenarios, an {\em energy-constrained scenario} with the mean of the harvested energy of $2$, and a {\em power-constrained scenario} with the mean of the harvested energy of $4$. In the energy-constrained scenario, the average harvested energy is much lower than the maximum transmission power and the energy schedule is mainly constrained by the energy availability. On the other hand, in the power-constrained scenario, the average harvested energy approaches to the maximum transmission power and this constraint dominates the energy scheduling. For both scenarios, we compare the performance of the proposed algorithm with the standard discrete MDP method,  the greedy policy and the balanced policy, averaged over $2\times 10^6$ realizations in Fig.~\ref{fg:fint_ec} and Fig.~\ref{fg:fint_pc}, respectively.  Although we cannot obtain the optimal performance, we utilize the error bound given in \eqref{eq:bound} and \eqref{eq:errbound} as an upper-bound of the optimal performance. Also, the performance obtained by the standard discrete MDP method can serve as the lower-bound.

%
%

It can be seen from Fig.~\ref{fg:fint_ec} and Fig.~\ref{fg:fint_pc} that for $\delta=1$, the performance of the proposed algorithm tightly approaches the upper-bound of the optimal performance in both scenarios while there is a gap between the proposed algorithm and the standard discrete MDP method. It is mainly because that the discrete MDP method discretizes all continuous variables and causes some non-negligible error with the large discretization step. For $\delta = 0.1$,  both the proposed algorithm and the standard discrete MDP method achieve the comparable performance, but their computational complexities are not comparable, e.g., the exhaustive search is involved in the latter. The greedy and balanced policies both have significantly inferior performances. Moreover, we note that the total rate increases as the Rayleigh parameter $\sigma$ increases and the rate in the energy-constrained scenario is higher than that in the power-constrained scenario.

\begin{figure}
\centering
\includegraphics[width=0.75\textwidth]{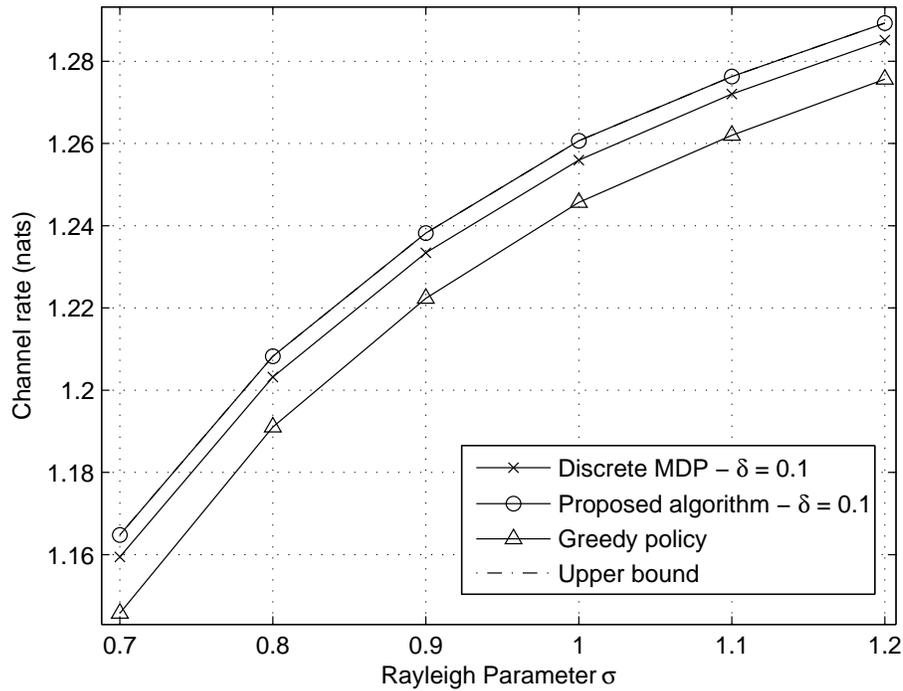}
\caption{Performance comparisons for the infinite-horizon case.}
\label{fg:inf}
\end{figure}

\begin{figure}
\centering
\includegraphics[width=0.75\textwidth]{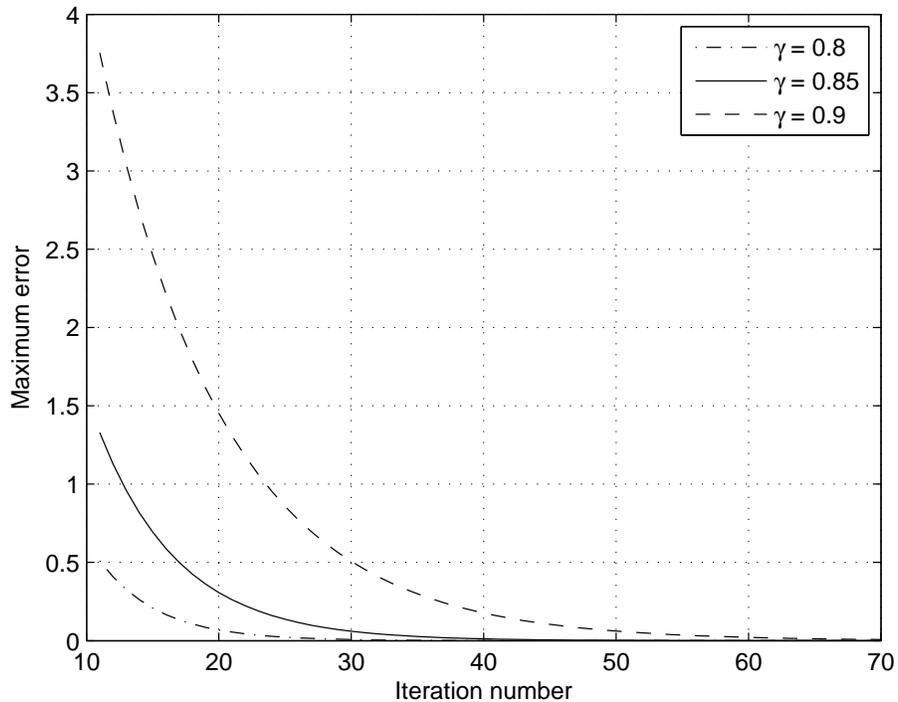}
\caption{The convergence of Algorithm 2 for $\sigma=1$.}
\label{fg:conv}
\end{figure}

For the infinite-horizon case, we set $\gamma=0.85$, $e_k = 3$, and $\sigma=0.7,0.8,0.9,1.0,1.1,1.2$.  Similar to the finite-horizon case, we evaluate the performance for various power allocation policies, averaged over $2\times10^6$ realizations. The performance comparisons for various power allocation policies are shown in Fig.~\ref{fg:inf}. Moreover, the convergence behavior of the proposed algorithm is also shown in Fig.~\ref{fg:conv} for $\sigma=1$ and $\gamma=0.8,0.85,0.9$.

Similar to the finite-horizon case, it is seen from Fig.~\ref{fg:inf} that the proposed algorithm has the best performance, tightly approaching the upper-bound of the optimal performance. We note that the  standard discrete MDP method with a discretization step of $\delta=0.1$ performs worse than  the  proposed algorithm. Further, the approximation gap is slightly higher in the infinite case as compared to that in the finite case. Moreover, we see that the greedy approach has the worst performance. In addition, it is seen from Fig.~\ref{fg:conv} that the discount factor affects the convergence speed, as analyzed in Section IV. Also, in the simulations for  $\gamma=0.8,0.85,0.9$, the proposed algorithm converges within around $30$, $40$ and $70$ iterations, respectively.

\begin{figure}[!htb]
\centering
\includegraphics[width=0.75\textwidth]{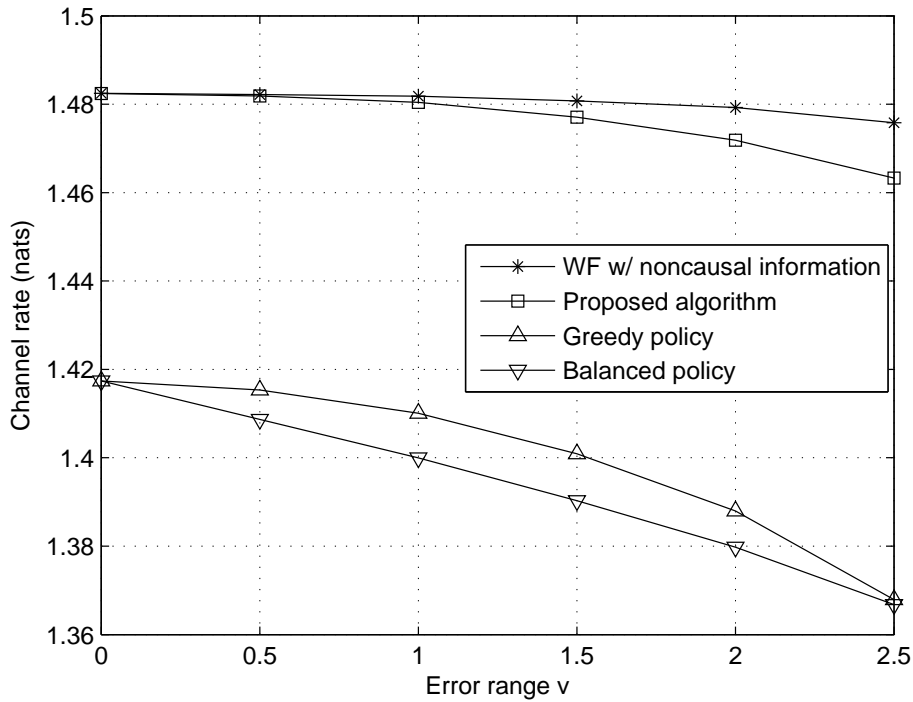}
\caption{Performance comparisons for the finite-horizon case with different prediction error ranges.}
\label{fg:var}
\end{figure}

We next evaluate the impact of the imperfect prediction error. We consider the finite-horizon case and set $K=10$, $\gamma=1$, $e_k = 3.5$, $\sigma=1$, $q=1-\hat{q}=0$, and $\delta = 0.1$. In this scenario, we only consider the impact of the imperfect prediction and we assume that the channel fading is known and the energy prediction error follows the discrete uniform distribution between $-v$ and $v$ with the step of $0.1$. The total payoff obtained by the proposed algorithm with causal information and the water-filling based algorithm in \cite{MyPaper} with non-causal information is compared in Fig. \ref{fg:var}, over different prediction error  ranges $v=0,0.5,1,1.5,2,2.5$. It is seen from Fig. \ref{fg:var} that as  $v$ decreases, the performance gap of the two algorithms with and without non-causal information decreases and approaches zero.

\section{Conclusions}
We have considered the problem of optimal power allocation for an  access-controlled transmitter with energy harvesting capability, operating in time-slotted fashion with causal knowledge of the channel state and the energy harvesting state. The energy harvesting process is a sum of a deterministic non-causal estimate and a random causal prediction error. This problem is formulated as a Markov decision process with continuous state. To efficiently solve this problem for both the finite- and infinite-horizon cases, we have introduced the approximate value function and developed efficient algorithms for obtaining the approximately optimal solutions.  The proposed algorithms provide an approximately optimal continuous power allocation, whose performance is better than that obtained by the standard discrete MDP method, in a computationally efficient manner. Simulation results demonstrate that the proposed algorithms can closely approach the optimal performance for both the finite- and infinite-horizon cases.

\appendices
\section{Proof of Lemma \ref{lm:cond}}
It is known that ${\cal T}_{0}$, which is the operator in the standard value iteration algorithm, is a $\gamma$-contraction \cite{MDP}. Denoting $(b^*,A^*)\triangleq\arg||{\cal T}_{0}\Big[V_1(b,A)\Big] - {\cal T}_{0}\Big[V_2(b,A)\Big]||_{\infty}$, for any $(B_0,A_0)$ and $(B_0+\delta,A_0)$ where $B_0,B_0+\delta\in{\cal B}_{\delta},A_0\in\{0,1\}$, we have that
\begin{align}
&\left|{\cal T}_{0}\Big[V_1\Big](B_0,A_0) - {\cal T}_{0}\Big[V_2\Big](B_0,A_0)\right|\leq \left|\Big({\cal T}_{0}\Big[V_1\Big] - {\cal T}_{0}\Big[V_2\Big]\Big)(b^*,A^*)\right|\label{eq:001}\
\end{align}
and
\begin{align}
&\left|{\cal T}_{0}\Big[V_1\Big](B_0+\delta,A_0) - {\cal T}_{0}\Big[V_2\Big](B_0+\delta,A_0)\right|\leq \left|\Big({\cal T}_{0}\Big[V_1\Big] - {\cal T}_{0}\Big[V_2\Big]\Big)(b^*,A^*)\right|\label{eq:002}\ .
\end{align}

Note that, given a value function $V(b,A)$, ${\cal T}_{\delta}\Big[V\Big](b,A)$ is the piecewise linear function reconstructed from the sample set $\{{\cal T}_{0}\Big[V\Big](B,A)\;|\;B\in{\cal B}_{\delta}\}$, as in \eqref{eq:avf}. Since $B_0,B_0+\delta\in{\cal B}_{\delta}$, then for any $b\in[B_0,B_0+\delta]$, we  have

%
%
%

\begin{align}
&\left|{\cal T}_{\delta}\Big[V_1\Big] (b,A)- {\cal T}_{\delta}\Big[V_2\Big](b,A)\right|\nonumber\\
&\leq \max\Big\{\left|{\cal T}_{\delta}\Big[V_1\Big] (B_0,A)- {\cal T}_{\delta}\Big[V_2\Big](B_0,A)\right|,\left|{\cal T}_{\delta}\Big[V_1\Big] (B_0+\delta,A)- {\cal T}_{\delta}\Big[V_2\Big](B_0+\delta,A)\right|\Big\}\nonumber\\
&= \max\Big\{\left|{\cal T}_{0}\Big[V_1\Big](B_0,A) - {\cal T}_{0}\Big[V_2\Big](B_0,A)\right|,\left|{\cal T}_{0}\Big[V_1\Big](B_0+\delta,A) - {\cal T}_{0}\Big[V_2\Big](B_0+\delta,A)\right|\Big\}\label{eq:011}
\end{align}

Since $B_0$ and $A$ are arbitrarily chosen from ${\cal B}_{\delta}/\max\{{\cal B}_{\delta}\}$ and $\{0,1\}$, respectively, we have
\begin{align}
&||{\cal T}_{\delta}\Big[V_1(b,A)\Big] - {\cal T}_{\delta}\Big[V_2(b,A)\Big]||\nonumber\\ \leq& \max\Big\{\left|{\cal T}_{0}\Big[V_1\Big](B_0,A_0) - {\cal T}_{0}\Big[V_2\Big](B_0,A_0)\right|,\left|{\cal T}_{0}\Big[V_1\Big](B_0+\delta,A_0) - {\cal T}_{0}\Big[V_2\Big](B_0+\delta,A_0)\right|\Big\}\label{eq:021}\\
\leq &\left |\Big({\cal T}_{0}\Big[V_1\Big]-{\cal T}_{0}\Big[V_2\Big]\Big)(b^*,A^*)\right|\label{eq:022}\\
=& ||{\cal T}_{\delta}\Big[V_1(b,A)\Big] - {\cal T}_{\delta}\Big[V_2(b,A)\Big]||_{\infty}\label{eq:023}\\
\leq& \gamma||V_1(b,A)-V_2(b,A)||_{\infty}
\end{align}
where \eqref{eq:021} follows from \eqref{eq:011}, \eqref{eq:022} follows from \eqref{eq:001}-\eqref{eq:002}, and \eqref{eq:023} follows the definition of $(b^*,A^*)$.


\section{Proof of Theorem \ref{thm:inferr}}
Denote $\beta(b,A) \triangleq  v^*(b,A) - {\cal T}_{\delta}\Big[v^*(b,A)\Big]$.
By Lemma \ref{lm:cond}, we have
\begin{align}
&||W_{\delta}^{(i)}(b,A)-v^*(b,A)||_{\infty}\nonumber \\
&= ||W_{\delta}^{(i)}(b,A) + W_{\delta}^{(i+1)}(b,A)-W_{\delta}^{(i+1)}(b,A) - v^*(b,A)||_{\infty}\nonumber\\
&\leq ||W_{\delta}^{(i)}(b,A) - W_{\delta}^{(i+1)}(b,A)||_{\infty}+ ||W_{\delta}^{(i+1)}(b,A) - v^*(b,A)||_{\infty}\nonumber\\
&= ||{\cal T}_{\delta}\Big[ W_{\delta}^{(i)}(b,A)\Big] - {\cal T}_{\delta} \Big[W_{\delta}^{(i-1)}(b,A)\Big]||_{\infty}+ ||{\cal T}_{\delta}\Big[W_{\delta}^{(i)}(b,A)\Big] -  {\cal T}_{\delta}\Big[v^*(b,A)\Big] - \beta(b,A)||_{\infty}\nonumber\\
&= ||{\cal T}_{\delta} \Big[W_{\delta}^{(i)}(b,A)\Big] - {\cal T}_{\delta} \Big[W_{\delta}^{(i-1)}(b,A)\Big]||_{\infty}+ ||{\cal T}_{\delta}\Big[W_{\delta}^{(i)}(b,A)\Big] -  {\cal T}_{\delta}\Big[v^*(b,A)\Big]||_{\infty} + ||\beta(b,A)||_{\infty}\nonumber\\
&\leq \gamma  || W_{\delta}^{(i)}(b,A) -  W_{\delta}^{(i-1)}(b,A)||_{\infty}+ \gamma||W_{\delta}^{(i)}(b,A) +  v^*(b,A)||_{\infty}+ ||\beta(b,A)||_{\infty}\label{eq:3}
\end{align}
where \eqref{eq:3} follows the $\gamma$-contraction of the operator ${\cal T}_{\delta}$.

From  \eqref{eq:3}, we have
\begin{align}
||W_{\delta}^{(i)}(b,A)-v^*(b,A)||_{\infty} & \leq\frac{\gamma||W_{\delta}^{(i)}(b,A) -  W_{\delta}^{(i-1)}(b,A)||_{\infty} + ||\beta(b,A)||_{\infty} }{1-\gamma}\nonumber\\& \leq \frac{\gamma\alpha + ||\beta(b,A)||_{\infty} }{1-\gamma} \label{eq:mvierr}
\end{align}
Also, since the only difference between $ {\cal T}_{\delta}$ and $ {\cal T}_{0}$ is the approximation process, then we have $\beta(b,A) = v^*(b,A) - {\cal T}_{\delta}\Big[v^*(b,A)\Big] =  v^*(b,A) - {\cal L}\Big[ {\cal T}_{0}\left[v^*(b,A)\right],\delta\Big]=  v^*(b,A) - {\cal L}\Big[v^*(b,A),\delta\Big]$. Using Proposition \ref{pp:err}, we have
\begin{equation}
||\beta(b,A)||_{\infty}\leq ||2v^*(\delta,A) - v(0,A) - v^*(2\delta,A)||_{\infty}\leq ||v^*(\delta,A) - v^*(0,A)||_{\infty} \ .
\end{equation}
Therefore, \eqref{eq:mvierr} can be further written as
\begin{equation}\label{eq:mvierr1}
||W_{\delta}^{(i)}(b)-v^*(b)||_{\infty} \leq \frac{\gamma\alpha + ||2v^*(\delta,A) - v(0,A) - v^*(2\delta,A)||_{\infty} }{1-\gamma}\ .
\end{equation}

%
%


%




\bibliographystyle{IEEETran}
\bibliography{IEEEabrv,bib}

\begin{thebibliography}{10}
\providecommand{\url}[1]{#1}
\csname url@samestyle\endcsname
\providecommand{\newblock}{\relax}
\providecommand{\bibinfo}[2]{#2}
\providecommand{\BIBentrySTDinterwordspacing}{\spaceskip=0pt\relax}
\providecommand{\BIBentryALTinterwordstretchfactor}{4}
\providecommand{\BIBentryALTinterwordspacing}{\spaceskip=\fontdimen2\font plus
\BIBentryALTinterwordstretchfactor\fontdimen3\font minus
  \fontdimen4\font\relax}
\providecommand{\BIBforeignlanguage}[2]{{%
\expandafter\ifx\csname l@#1\endcsname\relax
\typeout{** WARNING: IEEEtran.bst: No hyphenation pattern has been}%
\typeout{** loaded for the language `#1'. Using the pattern for}%
\typeout{** the default language instead.}%
\else
\language=\csname l@#1\endcsname
\fi
#2}}
\providecommand{\BIBdecl}{\relax}
\BIBdecl

\bibitem{GreenRadio}
C.~Han and et~al, ``Green radio: radio techniques to enable energy-efficient
  wireless networks,'' \emph{{IEEE} Commun. Mag.}, vol.~49, no.~6, pp. 46--54,
  Jun. 2011.

\bibitem{FundamentalTradeoff}
Y.~Chen and et~al, ``Fundamental trade-offs on green wireless networks,''
  \emph{{IEEE} Commun. Mag.}, vol.~49, no.~6, pp. 30--37, Jun. 2011.

\bibitem{Green4G}
S.~Yeh, ``Green {4G} communications: renewable-energy-based architectures and
  protocols,'' in \emph{Proc. 2010 Global Mobile Congress}, Oct. 2010, pp.
  1--5.

\bibitem{EnergyHarvesting}
S.~Sudevalayam and P.~Kulkarni, ``Energy harvesting sensor nodes: survey and
  implications,'' \emph{{IEEE} Commun. Surveys Tuts.}, vol.~13, no.~3, pp.
  443--461, Sep. 2011.

\bibitem{ASurvey}
A.~P. Bianzino and et~al, ``A survey of green networking research,''
  \emph{{IEEE} Commun. Surveys Tuts.}, vol.~14, no.~1, pp. 3--20, Jan. 2012.

\bibitem{FiniteHorizon}
S.~Chen, P.~Sinha, N.~Shroff, and C.~Joo, ``Finite-horizon energy allocation
  and routing scheme in rechargeable sensor networks,'' in \emph{Proc. IEEE
  2011 INFOCOM}, Apr. 2011, pp. 2273--2281.

\bibitem{OptimalEnergy}
C.~Ho and R.~Zhang, ``Optimal energy allocation for wireless communications
  with energy harvesting constraints,'' \emph{{IEEE} Trans. Signal Process.},
  vol.~60, no.~9, pp. 4808--4818, Sep. 2012.

\bibitem{TransmissionEnergy}
O.~Ozel, K.~Tutuncuoglu, J.~Yang, S.~Ulukus, and A.~Yener, ``Transmission with
  energy harvesting nodes in fading wireless channels: optimal policies,''
  \emph{{IEEE} J. Sel. Areas Commun.}, vol.~29, no.~8, pp. 1732--1743, Sep.
  2011.

\bibitem{MyPaper}
Z.~Wang, V.~Aggarwal, and X.~Wang, ``Iterative dynamic water-filing for fading
  multiple-access channels with energy harvesting,'' \emph{{IEEE} Trans. Signal
  Process.}, submitted for publication.

\bibitem{Neely}
L.~Huang and M.~Neely, ``Utility optimal scheduling in energy-harvesting
  networks,'' \emph{{IEEE/ACM} Trans. Netw.}, vol.~21, no.~4, pp. 1117--1130,
  Aug. 2013.

\bibitem{LearningTh}
P.~Blasco, D.~Gunduz, and M.~Dohler, ``A learning theoretic approach to energy
  harvesting communication system optimization,'' \emph{{IEEE} Trans. Wireless
  Commun.}, vol.~12, no.~4, pp. 1872--1882, Apr. 2013.

\bibitem{DiscretePI}
Q.~Bai, R.~Amjad, and J.~Nossek, ``Average throughput maximization for energy
  harvesting transmitters with causal energy arrival information,'' in
  \emph{Proc. IEEE 2013 WCNC}, Apr. 2013, pp. 4232--4237.

\bibitem{PMEHW}
J.~Piorno, C.~Bergonzini, K.~Atienza, and T.~Rosing, ``Prediction and
  management in energy harvested wireless sensor nodes,'' in \emph{Proc. VITAE
  2009}, May 2009, pp. 6--10.

\bibitem{AMPEAE}
J.~Lu, S.~Liu, Q.~Wu, and Q.~Qiu, ``Accurate modeling and prediction of energy
  availability in energy harvesting real-time embedded systems,'' in
  \emph{Proc. Green Computing Conf. 2010}, Aug. 2010, pp. 469--476.

\bibitem{MDP}
M.~Puterman, \emph{Markov Decision Processes: Discrete Stochastic Dynamic
  Programming}.\hskip 1em plus 0.5em minus 0.4em\relax New York: John Wiley \&
  Sons, 1994.

\bibitem{CO}
S.~Boyd and L.~Vandenberghe, \emph{Convex Optimization}.\hskip 1em plus 0.5em
  minus 0.4em\relax Cambridge: Cambridge University Press, 2009.

\end{thebibliography}
\end{document}